\documentclass[paper]{JHEP}
\usepackage{epsfig}
\def\MSbar{{$\overline{\mbox{\rm MS}}$}}
\def\half{{\textstyle {1\over2}}}
\def\beq{\begin{equation}}   \def\eeq{\end{equation}}
\def\beeq{\begin{eqnarray}}   \def\eeeq{\end{eqnarray}}
\def\as{\alpha_s}
\def\eps{\epsilon}
\def\ee{e^+e^-}
\def\epj#1#2#3{{\it Eur.\ Phys.\ J.\ }{\bf C#1} (#2) #3}
\def\spj#1#2#3{{\it Sov.\ Phys.\ JETP\ }{\bf #1} (#2) #3}
\def\cA{{\cal{A}}}
\def\cF{{\cal{F}}}
\def\cG{{\cal{G}}}
\def\cK{{\cal{K}}}
\def\aw{\bar\alpha_1}
\def\ae{\alpha_{\mbox{\scriptsize eff}}}
\def\aPT{\as^{\mbox{\scriptsize PT}}}
\def\mI{\mu_{\mbox{\tiny I}}}
\def\mR{\mu_{\mbox{\tiny R}}}
\def\re#1{(\ref{#1})}
\title{Power Corrections to Fragmentation Functions
in Non-Singlet Deep Inelastic Scattering}
\author{M.\ Dasgupta, G.E.\ Smye and B.R.\ Webber\thanks{Research
supported in part by the U.K. Particle Physics and Astronomy Research
Council and by the EC Programme ``Training and Mobility of Researchers",
Network ``Hadronic Physics with High Energy Electromagnetic Probes",
contract ERB FMRX-CT96-0008.}\\
        Cavendish Laboratory, University of Cambridge,\\
        Madingley Road, Cambridge CB3 0HE, U.K.\\
        E-mail: \email{dasgupta@hep.phy.cam.ac.uk}, etc.}
\abstract{We investigate the power-suppressed corrections to the
fragmentation functions of the current jet in non-singlet deep
inelastic lepton-hadron scattering. The current jet is defined
by selecting final-state particles in the current hemisphere
in the Breit frame of reference. Our method
is based on an analysis of one-loop Feynman graphs containing a
massive gluon, which is equivalent to the evaluation of leading 
infrared renormalon contributions. We find that the leading
corrections are proportional to $1/Q^2$, as in $\ee$ annihilation,
but their functional forms are different.
We give quantitative estimates based on the hypothesis
of universal low-energy behaviour of the strong coupling.}
\preprint{Cavendish--HEP--98/01}
\begin{document}
\section{Introduction}
The study of final-state properties in deep inelastic lepton
scattering (DIS) has received a great impetus from the increasing
quantity and kinematic range of the HERA data.  Amongst the
most interesting quantities being studied are the fragmentation
functions \cite{fragexp}, which specify the single-hadron momentum
distributions resulting from the fragmentation of the struck parton.
Although these functions cannot be calculated using perturbative QCD,
their asymptotic scaling violations (logarithmic $Q^2$ dependence)
can be predicted and used to measure the strong coupling $\as$,
in a similar way to the scaling violations in the deep inelastic
structure functions.  In addition, the fragmentation functions
and their scaling violations can be compared with those measured
in other processes, such as $\ee$ annihilation.

One problem with the measurement of $\as$ using scaling violation,
in either structure or fragmentation functions, is that there
is $Q^2$ dependence associated with power-suppressed (higher-twist)
contributions, in addition to the dominant logarithmic dependence.
These contributions need to be estimated in order to make use
of the wide $Q^2$ coverage of HERA.

Recently so-called `renormalon' or `dispersive' methods of
estimating power-suppressed terms have been suggested.
By looking at the
behaviour of the QCD perturbation series in high orders, one can
identify unsummable, factorially divergent sets of contributions
(infrared renormalons \cite{renormalons, BBNNA, Neu}) which indicate
that non-perturbative power-suppressed corrections must
be included. The $Q^2$-dependence of the leading correction
to a given quantity can be inferred, and by making further
universality assumptions one may also estimate its magnitude.
Tests of these ideas provide information on the transition
from the perturbative to the non-perturbative regime in QCD.
In particular, one can investigate the possibility that an
approximately universal low-energy form of the strong coupling
may be a useful phenomenological concept \cite{DokUra}--\cite{BPY}.

Such an approach has been applied to a wide variety
of observables, including DIS
structure functions \cite{BPY}--\cite{MeySchH1}, $\ee$ fragmentation
functions \cite{frag, BBMfrag}, and event shape variables in $\ee$
annihilation \cite{hadro}--\cite{Milan} and DIS \cite{DasWebshap, inprep}.
Comparisons with experimental data \cite{DELPHI}--\cite{KKPS} have
been encouraging.

Here we extend the same method to DIS fragmentation functions,
considering in the present paper the contribution which is
non-singlet with respect to the incoming hadron.  The
singlet part, which involves the gluon distribution function,
is formally non-leading in our approach, but may nevertheless
be important in the HERA kinematic region, where the gluon
density is high. Estimates of power corrections to singlet
structure functions have become available very
recently \cite{singlet}; we hope to apply similar
techniques to fragmentation functions in a future publication.
Meanwhile we concentrate on the non-singlet part, where
methods similar to those used for $\ee$ annihilation can be
applied. We find that the predicted leading power corrections
are proportional to $1/Q^2$, as in $\ee$ annihilation,
but their functional forms are different. The hypothesis
that power corrections are related to a universal low-energy
form for the strong coupling implies that their magnitudes
are given by a single non-perturbative parameter. We give
quantitative estimates based on the value of this parameter
derived from DIS structure function data \cite{DasWebDIS}.

In the following Section we review the approach of Ref.~\cite{BPY}.
Sect.~\ref{sec_frag} presents the standard leading-order
perturbative treatment of DIS fragmentation, which we modify
in Sect.~\ref{sec_pow} to estimate the non-singlet
power-suppressed corrections using the method
outlined in Sect.~\ref{sec_BPY}. Some numerical results and
conclusions are presented in Sect.~\ref{sec_conc}.

\section{Dispersive estimation of power corrections}\label{sec_BPY}
We assume that the QCD coupling $\as(k^2)$ can be defined down to
arbitrarily low values of the scale $k^2$ and that it has reasonable
analytic properties, i.e.\ no singularities other than a cut
along the negative real axis. It follows that one can write the
formal dispersion relation
\beq\label{alphas}
\as(k^2) = -\int_0^\infty \frac{d\mu^2}{\mu^2+k^2}\,\rho_s(\mu^2)
\eeq
where the `spectral function' $\rho_s$ represents the discontinuity
across the cut,
\beq\label{rhos}
\rho_s(\mu^2) = \frac{1}{2\pi i}\mbox{Disc}\left\{\as(-\mu^2)\right\}
\equiv \frac{1}{2\pi i}\left\{\as\left(\mu^2 e^{i\pi}\right)
-\as\left(\mu^2 e^{-i\pi}\right)\right\}\;.
\eeq

We now consider the calculation of some observable
$F$ in an ``improved one-loop'' approximation, \i.e.\  taking
into account one-gluon contributions plus those higher-order terms
that lead to the running of $\as$. As discussed in Ref.~\cite{BPY},
we expect that
\beq\label{Frel1}
F = \as(0)\cF(0) + \int_0^\infty \frac{d\mu^2}{\mu^2}\,\rho_s(\mu^2)
\,\cF(\mu^2/Q^2)
\eeq
where the {\it characteristic function} $\cF(\mu^2/Q^2)$ is obtained
by one-loop evaluation of $F$ (divided by $\as$) with the gluon
mass set equal to $\mu$ \cite{BBNNA, hadro}. The first term
on the right-hand side represents the contributions in which a single
gluon is produced or exchanged, while the second represents those with more
complex final or virtual states, e.g.\ the `decay products' of a virtual
gluon, which contribute to the running of $\as$. In contributions
that involve real multi-parton final states, Eq.~\re{Frel1} with the
full spectral function $\rho_s$ in the integrand is obtained only if
one sums inclusively over a sufficiently wide class of final states.
This point will be discussed more fully in Sect.~\ref{sec_pow}.

We can eliminate $\as(0)$ from Eq.~\re{Frel1} by means of the
dispersion relation \re{alphas}:
\beq\label{Frel2}
F = \int_0^\infty \frac{d\mu^2}{\mu^2}\,\rho_s(\mu^2)
\,\left[\cF(\mu^2/Q^2)-\cF(0)\right]\;.
\eeq

Non-perturbative effects at long distances are expected to give rise
to a modification in the strong coupling at low scales, $\delta\as$,
which generates a corresponding modification in the
spectral function via Eq.~\re{rhos}:
\beq\label{deltas}
\delta\rho_s(\mu^2)
= \frac{1}{2\pi i}\mbox{Disc}\left\{\delta\as(-\mu^2)\right\}\;.
\eeq

Inserting this in Eq.~\re{Frel2} and rotating the integration
contour separately in the two terms of the discontinuity, we obtain
the following non-perturbative contribution to the observable $F$:
\beq\label{deltaF}
\delta F = \int_0^\infty \frac{d\mu^2}{\mu^2}\,\delta\as(\mu^2)
\,\cG(\mu^2/Q^2)
\eeq
where, setting $\mu^2/Q^2 = \eps$,
\beq\label{Gdef}
\cG(\eps) = -\frac{1}{2\pi i}\mbox{Disc}\left\{\cF(-\eps)\right\}\;.
\eeq
Since $\delta\as(\mu^2)$ is limited to low values of
$\mu^2$, the asymptotic behaviour of $\delta F$ at large
$Q^2$ is controlled by the behaviour of $\cF(\eps)$ as
$\eps\to 0$.  We see from Eq.~\re{Gdef} that no terms analytic
at $\eps=0$ can contribute to $\delta F$.  On the
other hand for a square-root behaviour at small $\eps$,
\beq\label{deltaF1}
\cF\sim a_1 \frac{C_F}{2\pi}\sqrt{\eps}
\qquad\Longrightarrow\qquad
\delta F = -\frac{a_1}{\pi}\frac{\cA_1}{Q}\;,
\eeq
while
\beq\label{deltaF2}
\cF\sim a_2 \frac{C_F}{2\pi}\eps\ln\eps
\qquad\Longrightarrow\qquad
\delta F = a_2\frac{\cA_2}{Q^2}
\eeq
where
\beq\label{cAdef}
\cA_q \equiv \frac{C_F}{2\pi}
\int_0^\infty \frac{d\mu^2}{\mu^2}\,\mu^q\,\delta\as(\mu^2)\;.
\eeq

Notice that we express the result \re{deltaF} directly in
terms of the modification to the strong coupling itself,
rather than that in the derived quantity $\ae$ which was used
in some previous publications \cite{BPY, DasWebDIS, DokWeb97}:
\beq\label{deltaae}
\ae(\mu^2) =
\frac{\sin(\pi\mu^2\,d/d\mu^2)}{\pi\mu^2\,d/d\mu^2}\as(\mu^2)
= \as(\mu^2)-\frac{\pi^2}{6}\left(\mu^2\frac{d}{d\mu^2}\right)^2\as(\mu^2)
+\ldots\;.
\eeq
Although $\as$ and $\ae$ are similar in the perturbative region,
they differ substantially at low scales, and the former probably
has a simpler behaviour. For example, the even moments of the
effective coupling modification,
\beq\label{Adef}
A_{2p} \equiv \frac{C_F}{2\pi}
\int_0^\infty \frac{d\mu^2}{\mu^2}\,\mu^{2p}\,\delta\ae(\mu^2)\;,
\eeq
have to vanish for all integer values of $p$, whereas those of
$\delta\as$ do not. The translation dictionary for the moments
is in any case rather simple:
\beq\label{Arel}
\cA_{2p+1} = (-1)^p\,(p+\half)\pi\,A_{2p+1}\;,\;\;\;\;\;
\cA_{2p}   = (-1)^p\,p\,A'_{2p}\;,
\eeq 
where
\beq\label{Aprimedef}
A'_{2p} \equiv \frac{d}{dp}A_{2p} = \frac{C_F}{2\pi}
\int_0^\infty \frac{d\mu^2}{\mu^2}\,\mu^{2p}\,\ln\mu^2\,
\delta\ae(\mu^2)\;.
\eeq
Studies of power corrections to DIS structure functions
suggest that $\cA_2 = -A'_2\simeq 0.2$ GeV$^2$ \cite{DasWebDIS}.

As a clearer representation of the magnitudes of
power corrections, we may adopt the approach of
Refs.~\cite{DokWeb95, DokWeb97}
and express them directly in terms of moments of $\as$ over the
infrared region. We substitute for $\delta\as$ in Eq.~\re{deltaF}
\beq
\delta\as(\mu^2) \simeq \as(\mu^2) - \aPT(\mu^2)\;,
\eeq
where $\aPT$ represents the expression for $\as$
corresponding to the part already included in the
perturbative prediction.  As discussed in Ref.~\cite{DokWeb95},
if the perturbative calculation is carried out to second
order in the \MSbar\ renormalization scheme, with renormalization
scale $\mR^2$, then we have
\beq
\aPT(\mu^2) = \as(\mR^2) + [b\ln(\mR^2/\mu^2)+k]\,\as^2(\mR^2) 
\eeq
where for $N_f$ active flavours ($C_A=3$)
\beq
b = \frac{11 C_A-2N_f}{12\pi}\;,\;\;\;\;
k= \frac{(67-3\pi^2)C_A-10N_f}{36\pi}\;.
\eeq
The constant $k$ comes from a change of scheme from \MSbar\ to
the more physical scheme \cite{CMW} in which $\as$ is preferably
defined at low scales.
Then above some infrared matching scale $\mI$ we assume that
$\as(\mu^2)$ and $\aPT(\mu^2)$ approximately coincide, so that
\beeq\label{A2exp}
\cA_2 &\simeq& \frac{C_F}{2\pi}
\int_0^{\mI^2} d\mu^2\,\left(\as(\mu^2)-
\as(\mR^2) - [b\ln(\mR^2/\mu^2)+k]\,\as^2(\mR^2)\right)\nonumber\\
&=& \frac{C_F}{2\pi}\,\mI^2\,\left(\aw(\mI)-
\as(\mR^2) - [b\ln(\mR^2/\mI^2)+k+b]\,\as^2(\mR^2)\right)\;,
\eeeq
where
\beq
\aw(\mI) \equiv \frac{1}{\mI^2}\int_0^{\mI^2}\as(\mu^2)\,d\mu^2\;.
\eeq
The dependence of $\aw$ on $\mI$ is partially
compensated by the $\mI$-dependence of the other terms on
the right-hand side of Eq.~\re{A2exp}. The dependence on the
renormalization scale $\mR^2$ should help to compensate the
scale dependence of the perturbative part.  Notice that if
we take $\mR^2\propto Q^2$ then $\cA_2$ has a logarithmic dependence
on $Q^2$.  In general we do expect power corrections to have
additional logarithmic $Q^2$-dependences (anomalous dimensions),
but these are probably not given reliably by our `improved one-loop'
approximation.

\section{Fragmentation in DIS}\label{sec_frag}
A complication in DIS, absent from $\ee$ annihilation,
is the presence in the final state of the remnant of the
initial-state hadron, i.e.\ the constituents that did not
participate in the hard scattering of the lepton.  It is
expected that the fragmentation of the remnant will be
dominated by soft, non-perturbative physics. While of
interest for studying the hadronization process,
the remnant fragmentation is not so useful for QCD
test, and therefore we concentrate here on
aspects of fragmentation that are not sensitive to it.
This is conveniently done by looking at the final state
in the {\em Breit frame of reference} \cite{SWZ, PeRu, GDKT}.

We consider the deep inelastic scattering of a lepton of
momentum $l$ from a nucleon of momentum $P$, with momentum
transfer $q$.  The main kinematic variables are $Q^2 = -q^2$,  
the Bjorken variable $x=Q^2/2P\cdot q$ and
$y=P\cdot q/P\cdot l\simeq Q^2/x s$, $s$ being the total
c.m.\ energy squared. Then the Breit frame is the rest-frame
of $2xP+q$. In this frame the momentum transfer $q$ is purely
spacelike, and we choose to align it along the $+z$ axis:
\beq\label{Pqmoms}
P = \half Q (1/x,0,0,-1/x) \;,\;\;\;\;
q = \half Q (0,\,0,0,2)\;.
\eeq

To a good approximation, the fragmentation products of the remnant
will be moving in directions close to that of the incoming nucleon,
i.e.\ they will remain in the `remnant hemisphere' $H_r$ ($p_z<0$).
On the other hand the products of the hard lepton scattering
will tend to be found in the `current hemisphere' $H_c$
($p_z>0$). In fact in the parton model the scattered parton
moves along the current ($+z$) axis with momentum
$xP+q = \half Q (1,0,0,1)$.
Thus in the parton model the current hemisphere looks like
one hemisphere of the final state in $\ee$ annihilation at
centre-of-mass energy $Q$. Fragmentation studies have
shown that this similarity is indeed manifest in hadron
spectra and multiplicities \cite{fragexp}.  This makes it natural
to define the fragmentation functions in terms of particles $h$
appearing in the current hemisphere only, $h\in H_c$.

Since we wish to include only particles in the current hemisphere $H_c$,
we define the fragmentation function $F^h$ for a given hadron species
$h$ as a function of the variable $z= 2p_h\cdot q/ q^2$, which measures
the fraction of the hadron's momentum along the current direction and
takes values $0<z<1$ in $H_c$:
\beq\label{Dhdef}
F^h(z;x,Q^2) = \frac{d^3\sigma^h}{dx dQ^2 dz}
\bigg/ \frac{d^2\sigma}{dx dQ^2}\;. 
\eeq
The denominator of this expression is the fully inclusive deep inelastic
cross section,
\beq\label{DIScs}
\frac{d^2\sigma}{dx dQ^2} = \frac{2\pi\alpha^2}{Q^4}
\left\{\left[1+(1-y)^2\right] F_T(x)+2(1-y)F_L(x)\right\}
\eeq
where $F_T(x)=2F_1(x)$ and $F_L(x) = F_2(x)/x - 2F_1(x)$ are
the transverse and longitudinal structure functions (which also
have a weak $Q^2$-dependence that we do not show explicitly).
For simplicity we neglect here any contribution from weak interactions
(Z$^0$ or W$^\pm$ exchange). We shall comment on the effect of this
in Sect.~\ref{sec_conc}. The numerator in Eq.~\re{Dhdef}
is the single-hadron inclusive cross section,
\beq\label{diffcsh}
\frac{d^3\sigma^h}{dx dQ^2 dz} = \frac{2\pi\alpha^2}{Q^4}
\left\{\left[1+(1-y)^2\right] F^h_T(x,z)+2(1-y)
F^h_L(x,z)\right\}
\eeq
where $F^h_T$ and $F^h_L$ are generalized structure functions.

In the parton model (order $\as^0$), $F_L=F^h_L=0$ and
\beeq\label{Borncs}
F_T(x) &=& \sum_q e_q^2 [q(x)+\bar q(x)]\equiv f(x)\nonumber\\
F^h_T(x,z) &=& \sum_q e_q^2 [q(x)D_q(z)+\bar q(x)D_{\bar q}(z)]\;,
\eeeq
$q(x)$ and $\bar q(x)$ being the quark and antiquark momentum
fraction distributions in the target nucleon and
$D_q(z)$ and $D_{\bar q}(z)$ their fragmentation
functions for hadrons of type $h$.

\EPSFIGURE{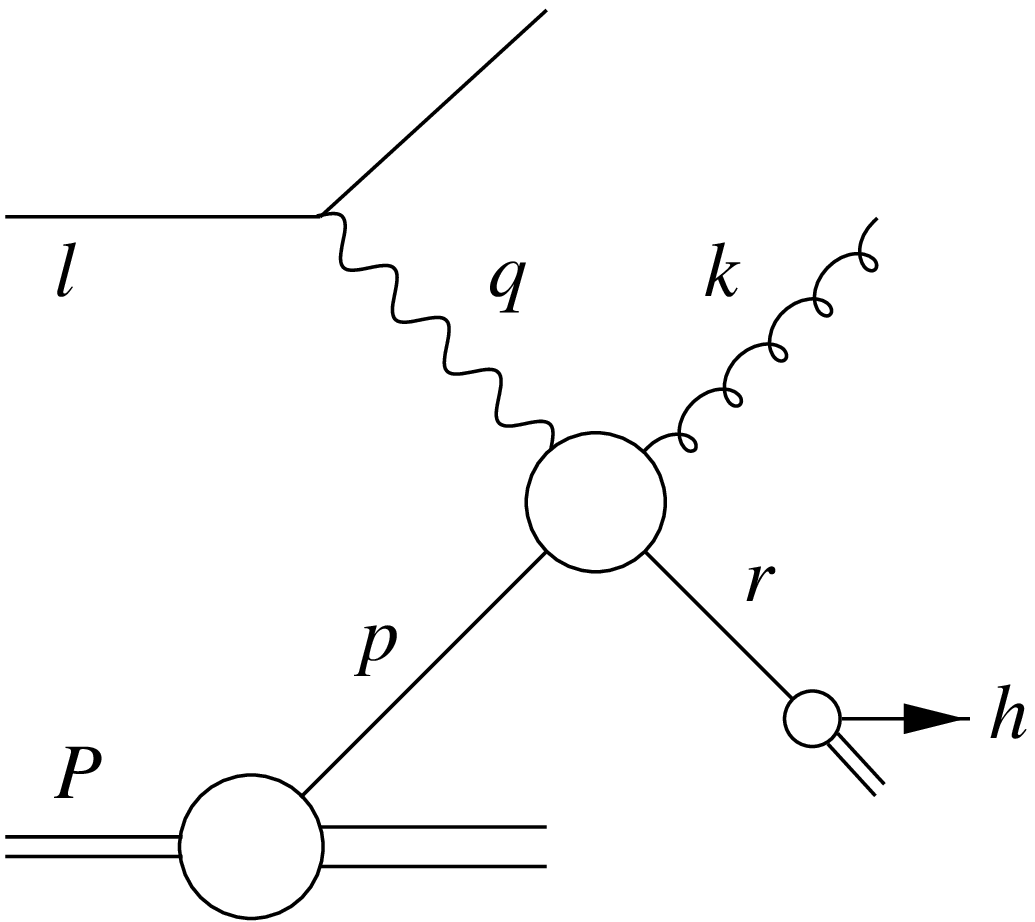}{Contribution to jet
fragmentation in deep inelastic scattering.}

To first order in $\as$, up to two final-state partons can be
emitted in the hard lepton-parton subprocess, as illustrated in
Fig.~1. The momentum of the struck parton is
$p = xP/\xi$ ($x<\xi<1$) and we define $\eta=P\cdot r/P\cdot q$ ($0<\eta<1$).
The parton-level cross section is 
\beq\label{diffcs}
\frac{d^3\sigma}{dx dQ^2 d\eta} = \frac{2\pi\alpha^2}{Q^4}
\left\{\left[1+(1-y)^2\right] F_T(x,\eta)+2(1-y) F_L(x,\eta)\right\}
\eeq
where (for $\eta<1$)
\beq
F_i(x,\eta) = \frac{\as}{2\pi}\sum_q e_q^2 \int_x^1 \frac{d\xi}{\xi}
\{C_F C_{i,q}(\xi,\eta) [q(x/\xi)+\bar q(x/\xi)]
+ T_R C_{i,g}(\xi,\eta)g(x/\xi)\}\;,
\eeq
$C_F=4/3$, $T_R=1/2$, and $g(x)$ is the gluon distribution. The coefficient
functions are \cite{PeRu}
\beeq\label{C0}
C_{T,q}(\xi,\eta) &=& \frac{\xi^2+\eta^2}{(1-\xi)(1-\eta)}
+2\xi \eta+2 \nonumber\\
C_{L,q}(\xi,\eta) &=& 4\xi \eta \nonumber\\
C_{T,g}(\xi,\eta) &=& \left[\xi^2+(1-\xi)^2\right]
\frac{\eta^2+(1-\eta)^2}{\eta(1-\eta)}\nonumber\\
C_{L,g}(\xi,\eta) &=& 8\xi(1-\xi)\;.
\eeeq

In the Breit frame $P$ and $q$ are given by Eq.~\re{Pqmoms}
and we can write
\beeq\label{moms}
p &=& \half Q (1/\xi,0,0,-1/\xi) \nonumber\\
r &=& \half Q (z_0,z_\perp,0,z_3) \nonumber\\
k &=& \half Q (\bar z_0,-z_\perp,0,\bar z_3)
\eeeq
where
\beeq\label{zs0}
z_0 &=& 2\eta-1+(1-\eta)/\xi\nonumber\\
z_3 &=& 1-(1-\eta)/\xi\nonumber\\
\bar z_0 &=& 1-2\eta+\eta/\xi\nonumber\\
\bar z_3 &=& 1-\eta/\xi\nonumber\\
z_\perp &=& 2\sqrt{\eta(1-\eta)(1-\xi)/\xi}\;.
\eeeq
We can distinguish four subregions of phase space,
as illustrated in Fig.~2:

A: both produced parton momenta $k,r$ in the current
hemisphere ($z_3,\bar z_3>0$);

B: only parton momentum $r$ in the current
hemisphere ($z_3 >0,\;\bar z_3<0$);

C: only parton momentum $k$ in the current
hemisphere ($z_3 <0,\;\bar z_3>0$);

D: no produced parton momenta in the current
hemisphere ($z_3, \bar z_3 <0$).

\EPSFIGURE{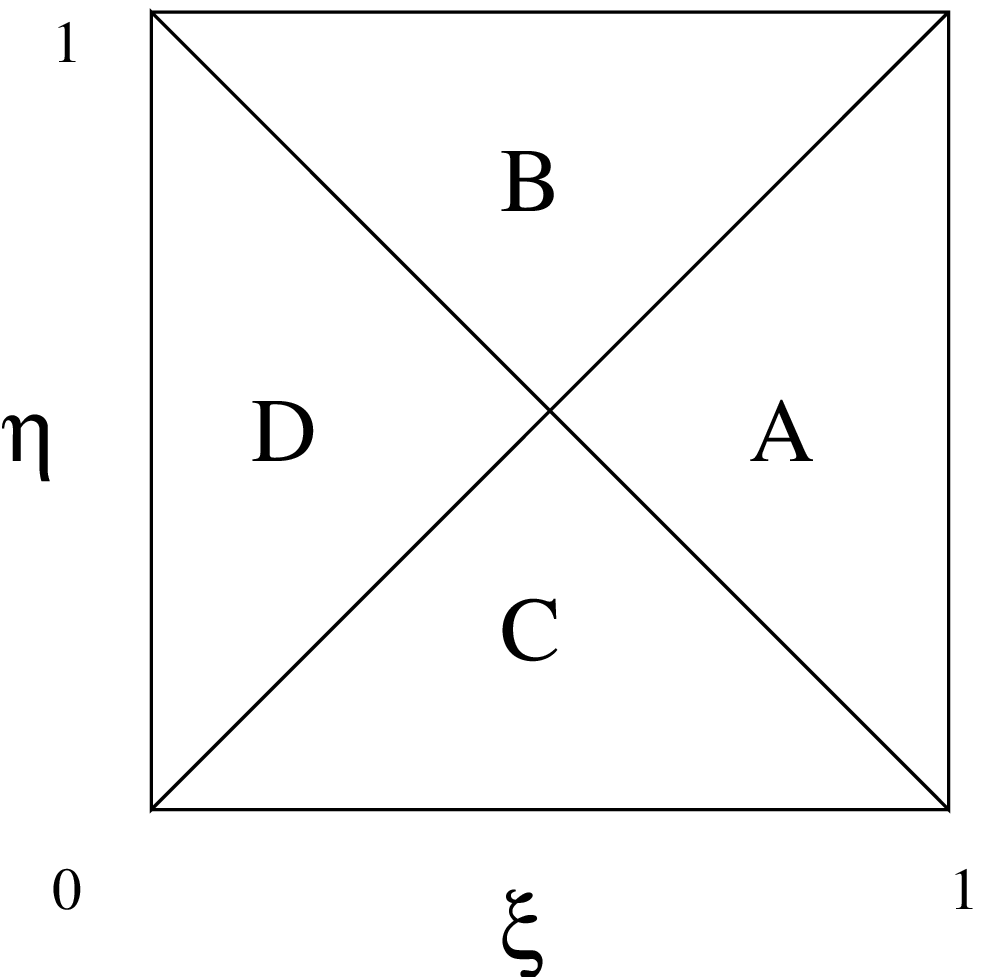}{Phase space region for
jet production in deep inelastic scattering.}

The  ${\cal O}(\as)$ contributions to the generalized structure functions
in Eq.~\re{diffcsh} are of the form
\beeq\label{FhOas}
F^h_i(x,z) &=& \frac{\as}{2\pi}\sum_q e_q^2
\int_x^1\frac{d\xi}{\xi}\int_0^1 d\eta\nonumber\\
&&\times\{C_F C_{i,q}(\xi,\eta) q(x/\xi)[\Theta(z_3)D_q(z/z_3)
+\Theta(\bar z_3)D_g(z/\bar z_3)]\nonumber\\
&&+\;C_F C_{i,q}(\xi,\eta) \bar q(x/\xi)[\Theta(z_3)D_{\bar q}(z/z_3)
+\Theta(\bar z_3)D_g(z/\bar z_3)]\nonumber\\
&&+\;T_R C_{i,g}(\xi,\eta)g(x/\xi)[\Theta(z_3)D_q(z/z_3)
+\Theta(\bar z_3)D_{\bar q}(z/\bar z_3)]\}
\eeeq
where $D_g$ is the gluon fragmentation function. In Eq.~\re{FhOas}
the coefficient functions $C_{i,p}$ include virtual corrections
at $\eta=1$, not shown in Eqs.~\re{C0}.

Changing the variable of integration from $\eta$ to $\zeta=z_3$ or
$\bar z_3$ as appropriate in each term, we can rewrite this as
\beeq\label{FhK}
F^h_i(x,z) &=& \sum_q e_q^2
\int_x^1 d\xi \int_z^1 d\zeta\nonumber\\
&&\times [K_{i,qq}(\xi,\zeta) q(x/\xi)D_q(z/\zeta)
        +K_{i,qg}(\xi,\zeta) q(x/\xi)D_g(z/\zeta)\nonumber\\
&&+\;K_{i,qq}(\xi,\zeta) \bar q(x/\xi)D_{\bar q}(z/\zeta)
   +K_{i,qg}(\xi,\zeta) \bar q(x/\xi)D_g(z/\zeta)\nonumber\\
&&+\;K_{i,gq}(\xi,\zeta) g(x/\xi)D_q(z/\zeta)
   +K_{i,gq}(\xi,\zeta) g(x/\xi)D_{\bar q}(z/\zeta)]
\eeeq
where
\beeq\label{Kqq}
K_{i,qq}(\xi,\zeta)&=&\frac{\as}{2\pi}
C_F C_{i,q}(\xi,1-\xi+\xi\zeta)\nonumber\\
K_{i,qg}(\xi,\zeta)&=&\frac{\as}{2\pi}
C_F C_{i,q}(\xi,\xi-\xi\zeta)\nonumber\\
K_{i,gq}(\xi,\zeta)&=&\frac{\as}{2\pi}
T_R C_{i,g}(\xi,\xi-\xi\zeta)\;.
\eeeq

It is sometimes useful to convert the convolution integrals in
Eq.~\re{FhK} to simple products by means of a double Mellin
transformation,
\beq\label{FMellin}
\tilde F(N,M) = \int_0^1 dx \int_0^1 dz\,x^{N-1} z^{M-1} F(x,z)\;,
\eeq
so that
\beq\label{tildeFhK}
\tilde F^h_i = \sum_q e_q^2\left(
 \tilde K_{i,qq}\tilde q\tilde D_q
+\tilde K_{i,qg}\tilde q\tilde D_g
+\tilde K_{i,qq}\tilde{\bar q}\tilde D_{\bar q}
+\tilde K_{i,qg}\tilde{\bar q}\tilde D_g
+\tilde K_{i,gq}\tilde g\tilde D_q
+\tilde K_{i,gq}\tilde g\tilde D_{\bar q}\right)
\eeq
where (note the different powers)
\beq\label{KMellin}
\tilde K_{i,qq}(N,M) =
\int_0^1 d\xi \int_0^1 d\zeta\,\xi^N \zeta^M K_{i,qq}(\xi,\zeta)\;.
\eeq

\section{Power corrections}\label{sec_pow}
To estimate power corrections according to the method outlined in
Sect.~\ref{sec_BPY}, we must recalculate the relevant observables
using a non-zero gluon mass-squared $\mu^2 =\eps Q^2$, and then
extract the terms that are non-analytic at $\eps=0$. In
contributions involving an incoming gluon, the ``massive'' gluon
should be treated as an internal line of the singlet process
$\gamma^* q \to q' \bar q' q$ \cite{singlet}. Thus the cross
section involves two massive gluons and is formally beyond
the ``improved one-loop'' approximation that we are using.
Nevertheless such contributions may be important at small $x$,
where the gluon density is large, as is the case at HERA.
For the present we avoid these complications by considering
the non-singlet contribution. For this we require only
the quark coefficient functions, which become
\beeq\label{Ceps}
C_{T,q}(\xi,\eta,\eps) &=&
\frac{(1-\eta)(1-\xi)+2\xi \eta(1-\eta)^2-\xi\eps}{(1-\eta-\xi\eps)^2}
+\frac{2\xi \eta(1-\eps)}{(1-\eta-\xi\eps)(1-\xi)}\nonumber\\
&&+\frac{(1-\eta)(1-\xi)-\xi\eps}{(1-\xi)^2}\nonumber\\
C_{L,q}(\xi,\eta,\eps) &=&
\frac{4\xi \eta(1-\eta)^2}{(1-\eta-\xi\eps)^2}
\eeeq
in place of those given in Eq.~\re{C0}.
The kinematic variables that give the momenta according to
Eq.~\re{moms} are now
\beeq\label{zseps}
z_0 &=& 2\eta-1+(1-\eta)/\xi-\eps\nonumber\\
z_3 &=& 1-(1-\eta)/\xi+\eps\nonumber\\
\bar z_0 &=& 1-2\eta+\eta/\xi+\eps\nonumber\\
\bar z_3 &=& 1-\eta/\xi-\eps\nonumber\\
z_\perp &=& 2\sqrt{\eta(1-\eta)(1-\xi)/\xi-\eps \eta}
\;.
\eeeq

\EPSFIGURE{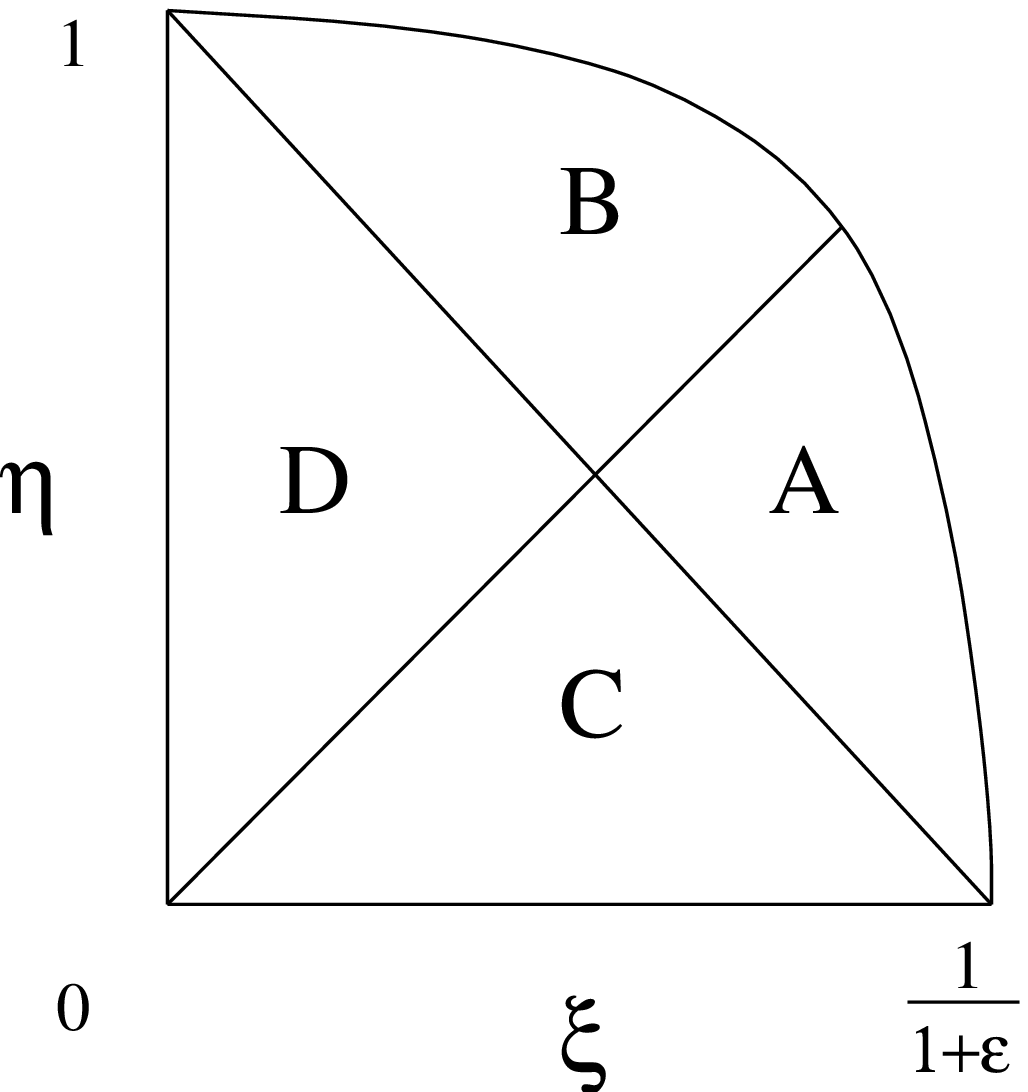}{Phase space region with
gluon mass-squared $\mu^2=\eps Q^2$.}

Thus the phase space region is $0 < \eta < 1-\eps \xi/(1-\xi)$, as
illustrated in Fig.~3. The regions A,\ldots D
defined above in terms of the signs of $z_3$ and $\bar z_3$ are
as indicated.

The corresponding characteristic functions for the
fragmentation functions are given by
Eq.~\re{FhK} with $K_{i,qq}(x,z)$ etc.\ replaced by
$(C_F/2\pi) \cK_{i,qq}(x,z,\eps)$ etc., where
\beeq\label{cFtau}
\cK_{i,qq}(\xi,\zeta,\eps) &=&\Theta(1-\xi-\eps \xi)\,
\Theta((1-\xi)(1-\zeta)-\eps \xi)\,
C_{i,q}(\xi,1-\xi+\xi \zeta-\eps \xi,\eps)\;,\nonumber\\
\cK_{i,qg}(\xi,\zeta,\eps) &=&\Theta(1-\xi-\eps \xi)\,
\Theta((1-\xi)(1-\xi+\xi \zeta)-\eps \xi^2)\,
C_{i,q}(\xi,\xi-\xi \zeta-\eps \xi,\eps)\;.
\eeeq
Note that for brevity we have extracted an overall factor of $C_F/2\pi$,
which will be absorbed in the non-perturbative factor \re{cAdef}.

Neglecting terms that are analytic or less singular than $\eps\ln\eps$
at $\eps=0$, we find that the Mellin transforms \re{KMellin} are given by
\beeq\label{tKeps}
\tilde\cK_{T,qq}(N,M,\eps)
&=&\left[2S_1(N)+ 2S_1(M+1)-3+\frac{1}{N}+\frac{1}{N+1}
+\frac{1}{M+1}+\frac{1}{M+2}\right]\ln\eps\nonumber\\ 
&&+\Biggl[-4S_1(N+1)-4S_1(M+1)+6+2N+2M+2NM \nonumber\\
&&-\frac{M+2}{N+1}-\frac{4}{N+2}-\frac{N+4}{M+1}
\Biggr]\eps\ln\eps\nonumber\\
\tilde\cK_{T,qg}(N,M,\eps)
&=& \left[-\frac{2}{M}+\frac{2}{M+1}
-\frac{1}{M+2}\right]\ln\eps + 
\left[1+\frac{4}{M}-\frac{N+4}{M+1}\right]\eps\ln\eps\nonumber\\
\tilde\cK_{L,qq}(N,M,\eps)
&=& \left[4-\frac{8}{N+2}\right]\eps\ln\eps\nonumber\\
\tilde\cK_{L,qg}(N,M,\eps) &=&0
\eeeq
where
\beq
S_1(N) = \sum_{j=1}^{N-1} \frac 1 j\;.
\eeq
Here we have included the virtual contribution, given in Ref.~\cite{BPY}.

The expression given above for the gluon fragmentation
contribution $\tilde\cK_{T,qg}$ is valid only for $M>2$.
There is an infrared divergence at $M=0$,
because we integrate the real gluon contribution
over one hemisphere only, which does not suffice to cancel
the divergent virtual contribution at $\zeta=0$.
For $M=1$ there is a contribution of $8\sqrt{\eps}$ instead
of an $\eps\ln\eps$ term, implying a $1/Q$-correction to this
moment of the gluon fragmentation function,
as is the case in $\ee$ annihilation \cite{frag, BBMfrag}.
For $M=2$ the 1 becomes --1 in the coefficient of $\eps\ln\eps$.
All of these changes represent extra contributions at the point
$\zeta=0$, which we can ignore because the fragmentation function
at any finite $z$ depends only on the behaviour at $\zeta>z>0$.

The $\ln\eps$ terms in Eqs.~\re{tKeps} generate the
logarithmic scaling violations in the structure and fragmentation
functions (see Ref.~\cite{BPY}), while the $\eps\ln\eps$
terms give rise to $1/Q^2$ power
corrections, as indicated in Eq.~\re{deltaF2}:
\beeq\label{deltaFh}
\delta F^h_i(x,z) &=&  \frac{\cA_2}{Q^2}
\sum_q e_q^2 \int_x^1 d\xi\int_z^1 d\zeta\nonumber\\
&&\times [H_{i,qq}(\xi,\zeta) q(x/\xi)D_q(z/\zeta)
        +H_{i,qg}(\xi,\zeta) q(x/\xi)D_g(z/\zeta)\nonumber\\
&&+\;H_{i,qq}(\xi,\zeta) \bar q(x/\xi)D_{\bar q}(z/\zeta)
   +H_{i,qg}(\xi,\zeta) \bar q(x/\xi)D_g(z/\zeta)]\;.
\eeeq
Inverting the Mellin transforms,  Eqs.~\re{tKeps} give
\beeq\label{HTexpn}
H_{T,qq}(\xi,\zeta)
&=& \left[\frac{4}{(1-\xi)_+}-2-4\xi-
+6\delta(1-\xi)+2\delta'(1-\xi)\right]\delta(1-\zeta)\nonumber\\
&&+ \left[\frac{4}{(1-\zeta)_+}-4
+2\delta'(1-\zeta)\right]\delta(1-\xi)\nonumber\\
&&-\delta'(1-\xi)-\delta'(1-\zeta)+2\delta'(1-\xi)\delta'(1-\zeta)\nonumber\\
H_{L,qq}(\xi,\zeta)
&=& \left[-8\xi+4\delta(1-\xi)\right]\delta(1-\zeta)\nonumber\\
H_{T,qg}(\xi,\zeta)
&=& \left[\frac{4}{\zeta}-4+\delta(1-\zeta)\right]\delta(1-\xi)
+\delta'(1-\xi)\;.
\eeeq

Eqs.~\re{deltaFh} and \re{HTexpn}, taken together with the value of
$\cA_2\simeq 0.2$ GeV$^2$ and the parton distribution and fragmentation
functions measured in other processes, provide a quantitative estimate
of the $1/Q^2$ corrections to current jet fragmentation in non-singlet
DIS.  Within the context of the dispersive method outlined in
Sect.~\ref{sec_BPY},
the contributions $H_{i,qq}$ from quark fragmentation should
be reliable, since they are integrated inclusively over
all other parton emission. The gluon contribution $H_{T,qg}$,
on the other hand, is less reliably estimated by the ``massive
gluon'' approach adopted here.  If, for example, a massive virtual
gluon splits into two partons, one of which goes into the remnant
hemisphere and one into the current hemisphere, then only the
latter will be counted and the full spectral function $\rho_s$
will not be built up in Eq.~\re{Frel1}. The limitations of the
massive gluon approach have been studied for $1/Q$ corrections
to event shapes in full two-loop order \cite{Milan, inprep},
and for $1/Q^2$ corrections to $\ee$ fragmentation functions
in the large-$N_f$ limit, i.e. including only quark
loops \cite{BBMfrag}. In the former case a significant but universal
enhancement of the power correction was found when going beyond the
massive gluon approximation, while in the latter case small corrections
were obtained, which could however become larger when gluon loops are
included. In both cases the massive gluon estimate of gluon
fragmentation effects provided a useful first approximation,
and we include it here in the same spirit.

\section{Results and conclusions}\label{sec_conc}
To obtain some indicative numerical results from
Eqs.~\re{deltaFh} and \re{HTexpn}, we assume for simplicity that
the quark fragmentation function $D_q$ is independent of quark flavour
and that $D_q=D_{\bar q}$. This is reasonable if heavy flavour
production is negligible and one sums over fragmentation into all
(charged) particles. Then, taking into account the parton-model
and non-perturbative contributions, we have from
Eqs.~\re{Dhdef}-\re{Borncs} and \re{deltaFh}
\beeq\label{Dhexpn}
F^h(z;x,Q^2) &=& D_q(z)+
\frac{\cA_2}{Q^2}\frac{1}{f(x)}
\int_x^1 d\xi \int_z^1 d\zeta\,f(x/\xi)\,
\{ [H_{T,qq}(\xi,\zeta)\nonumber\\
&&-H_{T,q}(\xi)\delta(1-\zeta)]D_q(z/\zeta)
 +H_{T,qg}(\xi,\zeta)D_g(z/\zeta) \}\;,
\eeeq
where $f(x)$ is the charge-weighted parton distribution given in
Eq.~\re{Borncs} and $H_{T,q}(\xi)$ is the
higher-twist coefficient function for the transverse structure
function \cite{BPY, DasWebDIS, Stein}\footnote{Note that the definition
here differs from that in Refs.~\cite{BPY, DasWebDIS, Stein} by a
factor of $-1/\xi$.}
\beq\label{HTq}
H_{T,q}(\xi) = \frac{4}{(1-\xi)_+}-2-4\xi
+4\delta(1-\xi)+\delta'(1-\xi)\;.
\eeq
It should be noted that the integral of $H_{T,qq}(\xi,\zeta)$
with respect to $\zeta$, over the range $0<\zeta<1$, is not
equal to $H_{T,q}(\xi)$, because the region $\zeta<0$ makes a
contribution of $2\delta(1-\xi)$ to the latter. On the other
hand the longitudinal contributions do satisfy the relation
\beq
H_{L,qq}(\xi,\zeta) = H_{L,q}(\xi)\delta(1-\zeta)
\eeq
and therefore no longitudinal higher-twist terms appear in
Eq.~\re{Dhdef}.
Inserting Eqs.~\re{HTexpn} and \re{HTq} into Eq.~\re{Dhexpn},
we find
\beq\label{Dhres}
F^h(z;x,Q^2) = D_q(z)\left(1+\frac{\cA_2}{Q^2}\,H(z;x)\right)
\eeq
where
\beeq\label{Dhfin}
H(z;x) &=&2xz\frac{f'(x)}{f(x)}\frac{D'_q(z)}{D_q(z)}
-x\frac{f'(x)}{f(x)}-2z\frac{D'_q(z)}{D_q(z)}
+\frac{D_g(z)}{D_q(z)}+2\nonumber\\
&&+\frac{4}{D_q(z)}\int_z^1 d\zeta\left[
\left(\frac{1}{(1-\zeta)_+}-1\right)D_q\left(\frac{z}{\zeta}\right)
+\left(\frac{1}{\zeta}-1\right)D_g\left(\frac{z}{\zeta}\right)\right]
\nonumber\\
&&+x\frac{f'(x)}{f(x)}\frac{1}{D_q(z)}\int_z^1 d\zeta
\left[D_q\left(\frac{z}{\zeta}\right)
     +D_g\left(\frac{z}{\zeta}\right)\right]\nonumber\\
&&+z\frac{D'_q(z)}{D_q(z)}\frac{1}{f(x)}
\int_x^1 d\xi\, f\left(\frac{x}{\xi}\right)\;.
\eeeq

\EPSFIGURE{fragcoef.eps}{Predicted coefficient of $\cA_2/Q^2$
for fragmentation function in non-singlet DIS. Dashed,
dot-dashed and solid curves are quark, gluon and total
fragmentation. The two sets of curves are for $x=0.3$ (upper)
and $0.1$ (lower).}

Figure~4 shows the resulting form of
$H(z;x)$ as a function of $z$ for various values of
$x$. We use the ALEPH \cite{ALEPH} parametrizations of the light quark
and gluon fragmentation functions for charged hadrons at $Q=22$ GeV, and the
corresponding MRST (central gluon) \cite{MRS98} parton distributions.
Thus the predictions are at $Q^2 = 484$ GeV$^2$, but $H(z;x)$ depends
only weakly (logarithmically) on $Q^2$, and in any case our method is
not reliable at the level of logarithmic variations.
Results become insensitive to $x$ below the values shown
in Fig.~4. Recall, however, that we have not computed the
singlet contribution, which may well be important at low $x$
because of the increase in the gluon distribution there. 

The predicted power corrections are qualitatively similar to those
for fragmentation functions in $\ee$ annihilation \cite{frag},
though somewhat larger in magnitude. Part of the increase
comes from the negative higher-twist correction to the transverse
structure function in the denominator of Eq.~\re{Dhdef}. The
rapid rise in the quark contribution at large $x$ and/or $z$
comes from the product of derivatives in the first term
of Eq.~\re{Dhfin}. The contribution from gluon fragmentation,
although subject to further corrections as discussed earlier,
is estimated to be relatively small for $z>0.2$.

Finally, we comment on the effect of taking weak interactions into
account. This introduces contributions from the parity-violating
structure function $F_3(x)$ in Eq.~\re{DIScs} and its analog $F^h_3(x,z)$
in Eq.~\re{diffcsh}. However, it was shown in Ref.~\cite{DasWebDIS} that
the predicted higher-twist contributions to $F_3$ and $F_T$ are the
same, i.e.\ $H_{3,q}= H_{T,q}$ as given in Eq.~\re{HTq}. Similarly
we find that $H_{3,qq}= H_{T,qq}$ and $H_{3,qg}= H_{T,qg}$ as given
in Eq.~\re{HTexpn}. Hence Eq.~\re{Dhfin} remains valid, provided the
parton distribution $f(x)$ is redefined with the appropriate
electroweak coefficients in place of the charges in Eq.~\re{Borncs}.

\acknowledgments
M.D.\ acknowledges the financial support of Trinity College, Cambridge.
B.R.W.\ thanks Yu.L.\ Dokshitzer for helpful comments.

\end{document}